\documentclass[epj]{webofc}
\usepackage[varg]{txfonts}   
\woctitle{MESON2016 - the 14$^\textrm{th}$ International Workshop on Meson 
	Production, Properties and Interaction}
\begin{document}
\selectlanguage{english}
\title{K-Long Facility for JLab and its Scientific Potential}
\author{Igor~I.~Strakovsky\inst{1}\fnsep\thanks{\email{igor@gwu.edu}} }
\institute{Institute for Nuclear Studies, Department of Physics,
        The George Washington University, Washington, D.C. 20052, U.S.A.}
\abstract{ 
Our main interest in creating a secondary high-quality KL-beam is to 
investigate hyperon spectroscopy through both formation and production 
processes. We propose to study two-body reactions induced by the 
KL-beam on the proton target. The experiment should measure both 
differential cross sections and self-analyzed polarizations of the 
produced $\Lambda$-, $\Sigma$-, and $\Xi$-hyperons using the GlueX 
detector at the Jefferson Lab Hall~D. New data will greatly constrain 
partial-wave analysis and reduce model-dependent uncertainties in the 
extraction of strange resonance properties, providing a new benchmark 
for comparisons with QCD-inspired models and LQCD calculations. The 
measurements will span c.m. $\cos\theta$ from -0.95 to 0.95 in c.m. 
range above W = 1490~MeV and up to 4000~MeV.}
\maketitle

\section{Introduction}
\label{intro}

At the beginning of February of 2016, Jefferson Lab hosted a 
Workshop \textit{Physics with Neutral Kaon beam at JLab}. It 
was dedicated to the physics of hyperons produced by the 
neutral Kaon beam on both unpolarized and polarized 
targets~\cite{KL2016}. The workshop follows our 
LoI--12--15--001~\cite{LOI2015} (JLab KLF Project) to help 
to address the comments made by the JLab PAC43 and to 
prepare the full proposal for PAC45. The emphasis is on the 
hyperon spectroscopy.  Mini-Proceedings of the KL2016 are 
available at arXiv~\cite{Proc2016}.  

The \textit{Excited Hyperons in QCD Thermodynamics at 
Freeze-Out} (YSTAR2016) Workshop~\cite{YSTAR2016} is a 
successor to the recent KL2016.  This workshop will discuss 
the influence of possible "missing" hyperon resonances on
QCD thermodynamics, on freeze-out in heavy ion collisions 
and in the early universe, and in spectroscopy. Recent 
studies that compare LQCD calculations of thermodynamic 
calculations, statistical hadron resonance gas models, and 
ratios between measured yields of different hadron species 
in heavy ion collisions provide indirect evidence for the 
presence of "missing" resonances in all of these contexts. 
The aim of the workshop is to sharpen these comparisons, 
advance our understanding of the formation of baryons from 
quarks and gluons microseconds after the Big Bang and in 
today's experiments, and to connect these developments to 
experimental searches for direct, spectroscopic, evidence 
for these resonances. 

The JLab12 energy upgrade, with the new Hall~D, is an ideal
tool for extensive studies of non-strange and, specifically,
strange baryon resonances~\cite{Curtis2016}.  Our plan is 
evolving to take advantage of the existing high quality 
photon beam line and experimental area in the Hall~D complex 
at Jefferson Lab to deliver a beam of $K_L$ particles
onto a liquid hydrogen cryotarget within the GlueX detector.  
The recently constructed GlueX detector in Hall~D
is a large acceptance spectrometer with good coverage for
both charged and neutral particles that can be adapted to
this purpose. Obviously, Kaon beam facility (KLF) with good
momentum resolution is crucial to provide the data needed to
identify and characterize the properties of hyperon
resonances. The masses and widths of the lowest $\Lambda$
and $\Sigma$ baryons were determined mainly with Kaon-beam
experiments in the 1970s~\cite{PDG2014}.  Pole position in 
complex energy plane for hyperons has began to be studied 
only recently, first of all for 
$\Lambda(1520)\frac{3}{2}^-$~\cite{Qiang2010}.

\section{Scope of the Proposed KLF Program}
\label{scope}

A comparison of recent coupled-channel analyses~\cite{Zhang2013,
Kamano2014,Jackson2015,Fernandez2016} comes to the conclusion 
that, for most cases, it is only the first excited state in 
each partial wave whose detailed properties [branching reaction 
(BRs), helicity amplitudes] are known. Different analyses may 
agree on the existence of the second state (in each partial 
wave) but not on their decay properties, while there is no 
agreement even on the existence of a third state in a 
particular partial wave. Given the arduous nature of the task 
involved in obtaining high-quality data that enter these 
multichannel analyses it is reasonable to address the question 
as to the final scope of this effort.  In other words: How many 
resonances do we need to identify in order to convince 
ourselves that we have achieved a solid understanding of the 
baryon spectrum from QCD? As examples, we examine this question 
from the viewpoint of lattice gauge and constituent quark model 
(QM) calculations.

Our current "experimental" knowledge of $\Lambda^\ast$,
$\Sigma^\ast$, $\Xi^\ast$, and $\Omega^\ast$ resonances is
far worse than our knowledge of $N^\ast$ and $\Delta^\ast$
resonances; though they are equally fundamental. Specifically, 
the properties of multi-strange baryons ($\Xi^\ast$ and 
$\Omega^\ast$ states) are poorly known. For instance the {\it 
Review of Particle Physics} lists only two states with BR to 
$K\Xi$, namely, $\Lambda(2100)\frac{7}{2}^-$ (BR $<$ 3\%) and
$\Sigma(2030)\frac{7}{2}^+$ (BR $<$ 2\%)~\cite{PDG2014}.
Clearly, complete understanding of three-quark bound states 
requires to learn more about baryon resonances in "strange 
sector" as well.

Reviewing analyses decades worth of data, from both hadronic 
and EM experiments, we have found numerous baryon resonances, 
and determined their masses, widths, and quantum numbers. There 
are 112 baryons in PDG2014 Listings~\cite{PDG2014} (including 
both non-strange and strange states) and only 58 of them are 
$4^\ast$ and $3^\ast$. Many more states have been predicted by 
QMs. For example in case of $SU(6)\times O(3)$, it would be 
required 434 resonances, if all revealed multiplets were 
completed (three $70^-$ and four $56^-$).

Three light quarks can be arranged in 6 baryonic families, 
$N^\ast$, $\Delta^\ast$, $\Lambda^\ast$, $\Sigma^\ast$, 
$\Xi^\ast$, and $\Omega^\ast$.  Number of members in a family 
that can exist is not arbitrary~\cite{Nefkens1997}.  If 
SU(3)$_F$ symmetry of QCD is controlling, then for the octet: 
$N^\ast$, $\Lambda^\ast$, and $\Sigma^\ast$, and for the 
decuplet: $\Delta^\ast$, $\Sigma^\ast$, $\Xi^\ast$, and 
$\Omega^\ast$.  Number of experimentally identified resonances 
of each baryon family in PDG2014 summary tables is 17 $N^\ast$, 
24 $\Delta^\ast$, 14 $\Lambda^\ast$, 12 $\Sigma^\ast$, 7 
$\Xi^\ast$, and 2 $\Omega^\ast$.  Constituent QMs, for 
instance, predict existence of no less than 64 $N^\ast$ and 22
$\Delta^\ast$ states with mass less than 3~GeV.  Seriousness 
of "missing-states" problem~\cite{Koniuk1980} is obvious from
these numbers. To complete SU(3)$_F$ multiplets, one needs no
less than 17 $\Lambda^\ast$, 41 $\Sigma^\ast$, 41 $\Xi^\ast$,
and 24 $\Omega^\ast$.

\section{Observables}
\label{Observables}

\begin{figure}[ht]
\begin{center}
\includegraphics[angle=90, width=0.35\textwidth ]{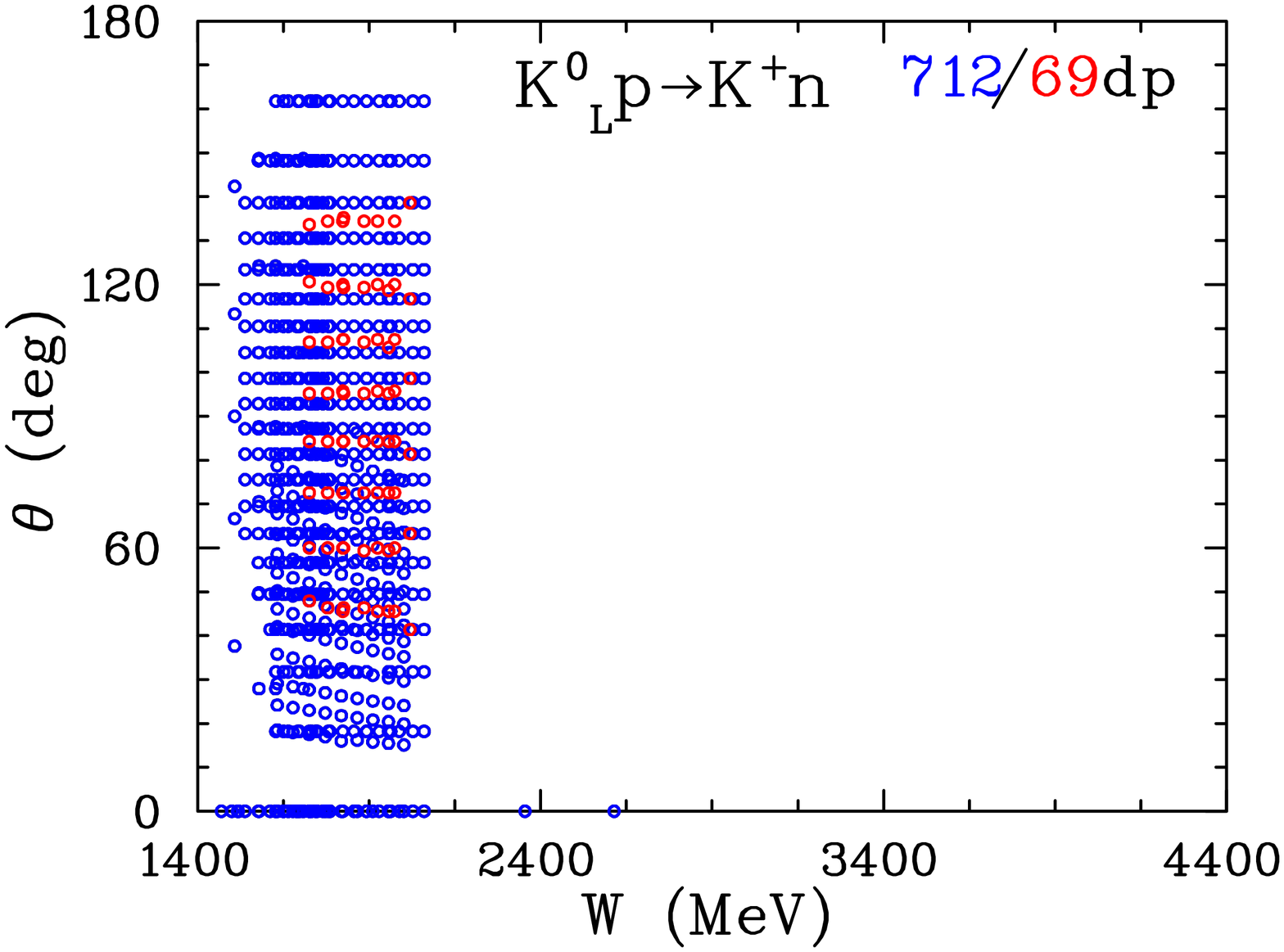}
\includegraphics[angle=90, width=0.35\textwidth ]{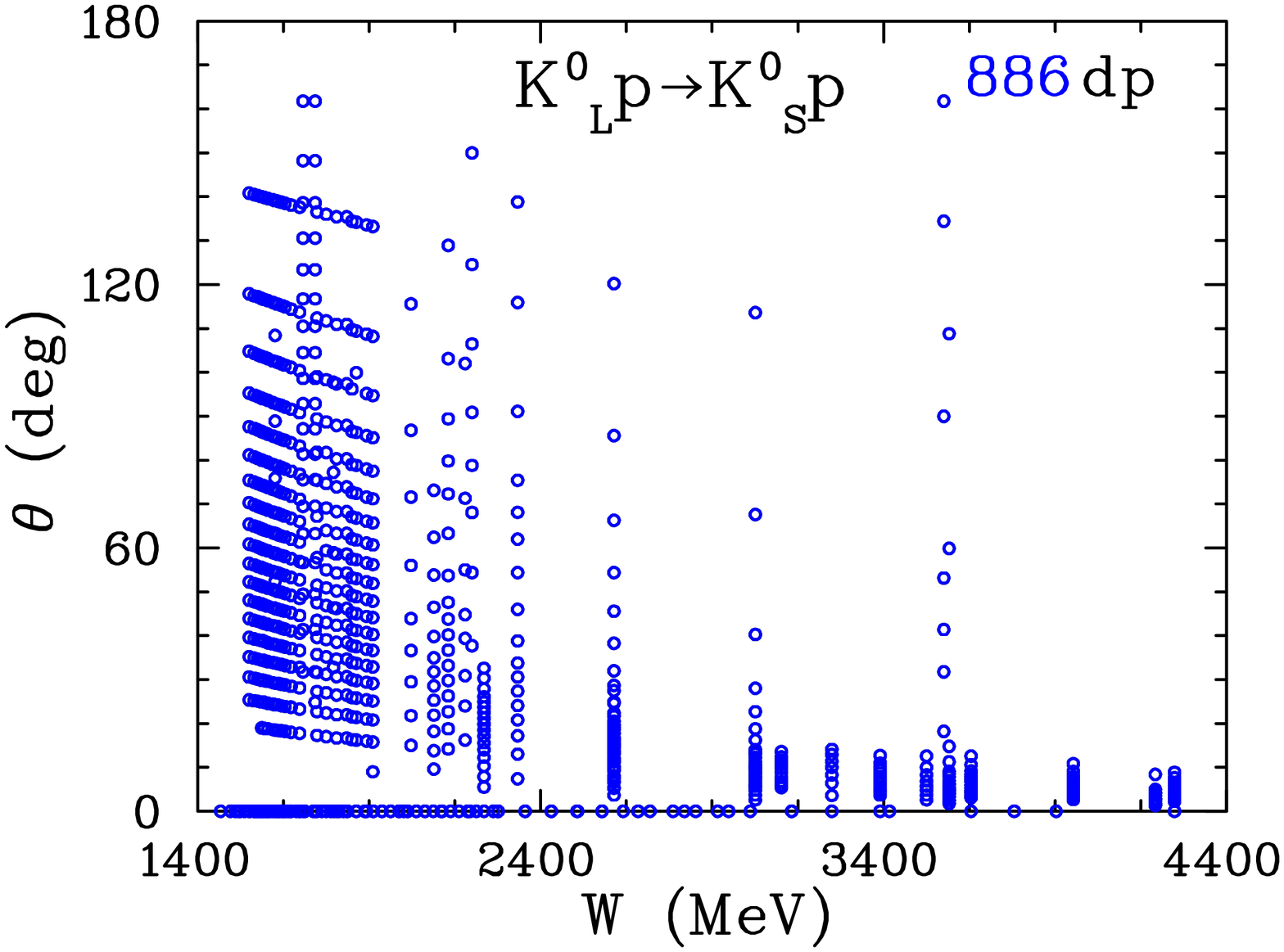}
\includegraphics[angle=90, width=0.35\textwidth ]{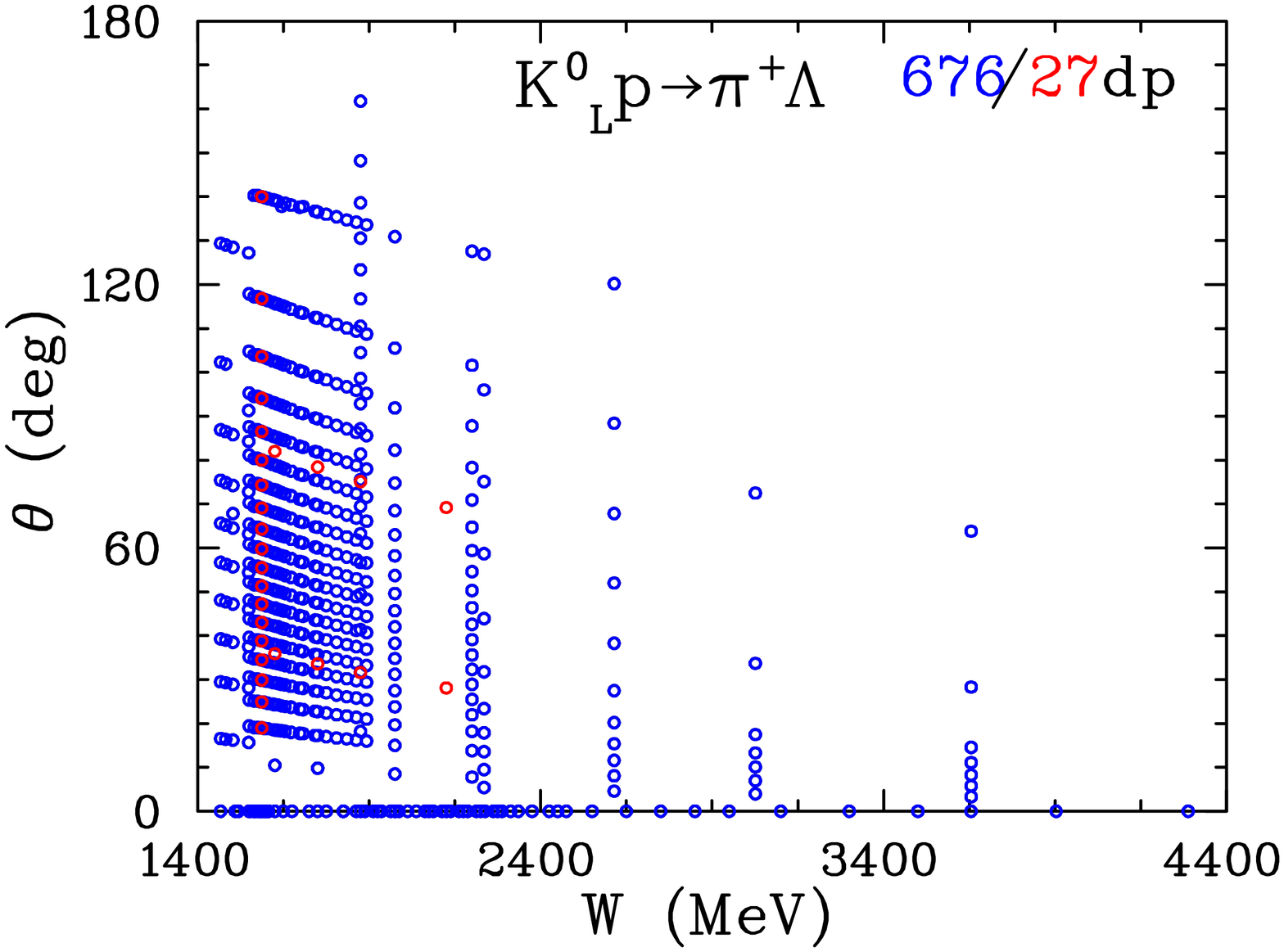}
\includegraphics[angle=90, width=0.35\textwidth ]{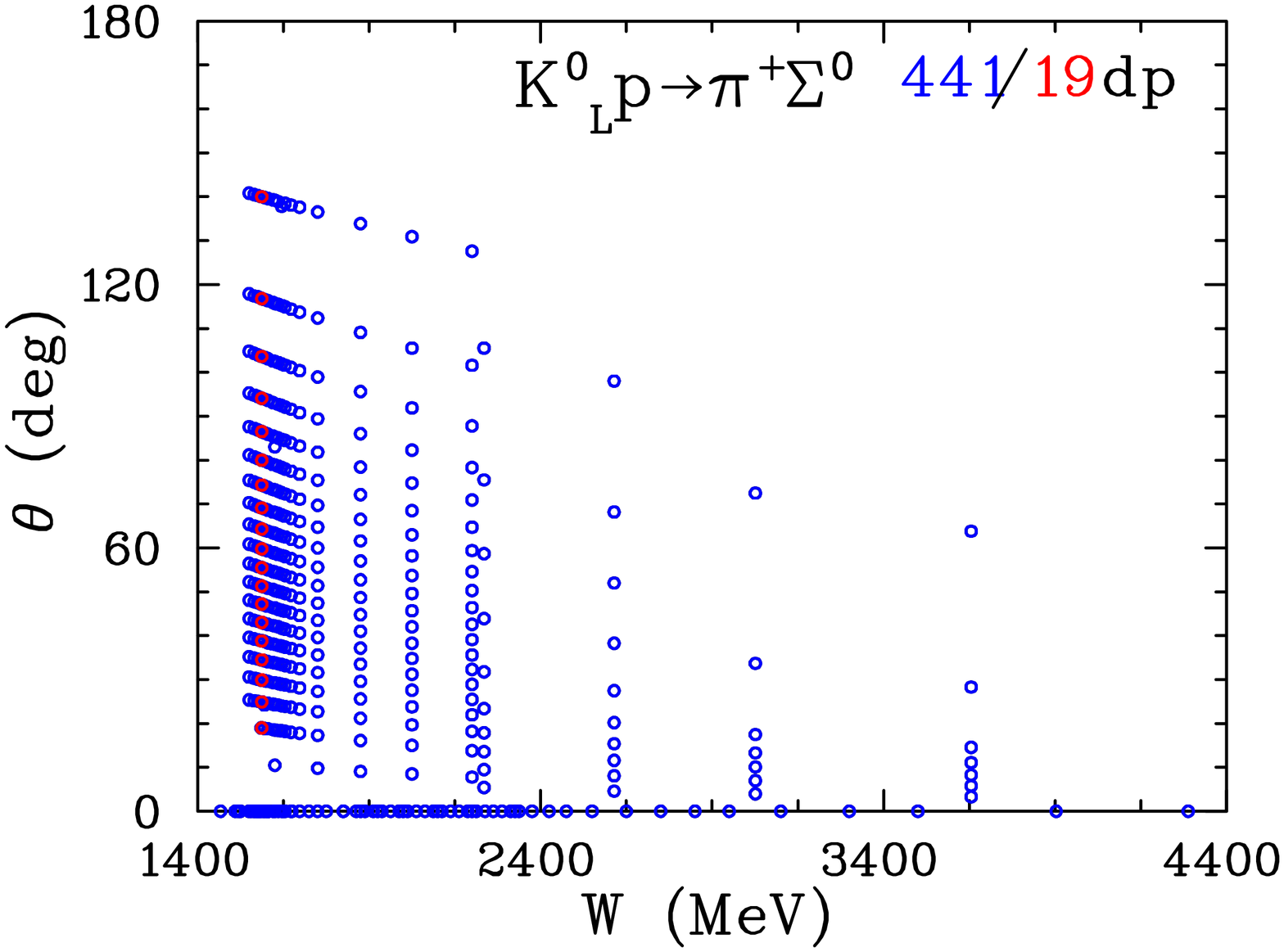}
\end{center}
\centerline{\parbox{0.80\textwidth}{
 \caption[] {\protect\small Experimental data available for
        $K_Lp\rightarrow K^+n$, 
        $K_Lp\rightarrow K_Sp$, 
	$K_Lp\rightarrow\pi^+\Lambda$, and
        $K_Lp\rightarrow\pi^+\Sigma^0$
	as a function of c.m. energy
        $W$~\protect\cite{Brem}. The number of
        data points (dp) is given in the upper righthand
        side of each subplot [blue (red) shows amount of
        unpolarized (polarized) observables]. Total cross
        sections are plotted at zero degrees.}
        \label{fig:fig1}}}
\end{figure}

There are two particles in the reactions $K_Lp\rightarrow\pi Y$
and $KY$ that can carry polarization: the target and recoil
nucleon/hyperon. Hence, there are two possible
double-polarization experiments: target/recoil. 
While a formally complete experiment requires the measurement,
at each energy and angle, of at least three independent
observables, the current database for $K_Lp\rightarrow\pi Y$
and $ KY$ is populated mainly by unpolarized cross sections.
Figure~\ref{fig:fig1} illustrates this quite clearly.

The experiments using unpolarized LD$_2$ (to get "neutron"
data) and polarized target (aka FROST) for both hydrogen and
deuteriun components, we will leave for the following
proposals.  Obviosly, it will open up a new avenue to the 
complete experiment. Note that the "neutron" data are 
critical to determine parameters of neutral $\Lambda^\ast$s
and $\Sigma^\ast$s hyperons which were considered 
recently~\cite{Zou2016}.

\section{Phenomenology / Partial-Wave Aanalysis}
\label{PWA}

Following H\"ohler~\cite{Hoehler1983}, the differential
cross section and polarization for $K_Lp\rightarrow\pi Y$ 
and $KY$ are given by
\begin{equation}
        \frac{d\sigma}{d\Omega} = \lambdabar^2 (|f(W,\theta)|^2 
	+ |g(W,\theta)|^2),~~~
        P\frac{d\sigma}{d\Omega} = 2\lambdabar^2 {\rm Im}
        (f(W,\theta)~g(W,\theta)^\ast) ,
\end{equation}
where $\lambdabar = \hbar/k$, with $k$ the magnitude of c.m.\ 
momentum for the incoming meson.  Here $f(W,\theta)$ and
$g(W,\theta)$ are the usual spin-nonflip and spin-flip
amplitudes at c.m.\ energy $W$ and meson c.m.\ scattering
angle $\theta$. In terms of partial waves, $f(W,\theta)$ and 
$g(W,\theta)$ can be expanded as
\begin{equation}
        f(W,\theta) = \sum_{l=0}^\infty [(l+1)T_{l+}
        + lT_{l-}]P_l(\cos\theta),~~~
        g(W,\theta) = \sum_{l=1}^\infty [T_{l+} - T_{l-}]P_l^1
        (\cos\theta),
\end{equation}
where $l$ is the initial orbital angular momentum,
P$_l$($\cos\theta$) is a Legendre polynomial, and
P$_l^1$($\cos\theta$) is an associated Legendre function.
The total angular momentum for the amplitude $T_{l+}$ is
$J=l+\frac{1}{2}$, while that for the amplitude $T_{l-}$
is $J=l-\frac{1}{2}$.  For hadronic scattering reactions,
we may ignore small CP-violating terms and write
\begin{equation}
        K_L^0 = \frac{1}{\sqrt{2}} (K^0 - \overline{K^0}),~~~
        K_S^0 = \frac{1}{\sqrt{2}} (K^0 + \overline{K^0}).
\end{equation}

We may generally have both $I=0$ and $I=1$ amplitudes for
$KN$ and $\overline{K}N$ scattering, so that the amplitudes
$T_{l\pm}$ can be expanded in terms of isospin amplitudes
as
\begin{equation}
        T_{l\pm} = C_0 T^0_{l\pm} + C_1 T^1_{l\pm},
\end{equation}
where $T_{l\pm}^I$ are partial-wave amplitudes with isospin
$I$ and total angular momentum $J = l\pm\frac{1}{2}$, with
the appropriate isospin Clebsch-Gordan coefficients $C_I$.

We plan to do a coupled-channel PWA with new GlueX KLF data in
combination with available and new J-PARC $K^-p$ measurements
when they will be available.  Then the best fit will allow to
determine data driven (model independent) partial-wave amplitudes
and associated resonance parameters as the SAID group does, for
instance, for analysis of $\pi$N-elastic, charge-exchange,
and $\pi^-p\rightarrow\eta n$ data~\cite{Arndt2006}.  With the
new GlueX KLF data, the quantitative significance of resonance
signals can be determined.  Additionally, new PWA with new GlueX
data will allow to look for "missing" hyperons via looking for new
poles in complex plane positions.  It will provide a new
benchmark for comparisons with QCD-inspired models and LQCD
calculations.
\begin{figure}[ht]
\begin{center}
\includegraphics[angle=90, width=0.35\textwidth]{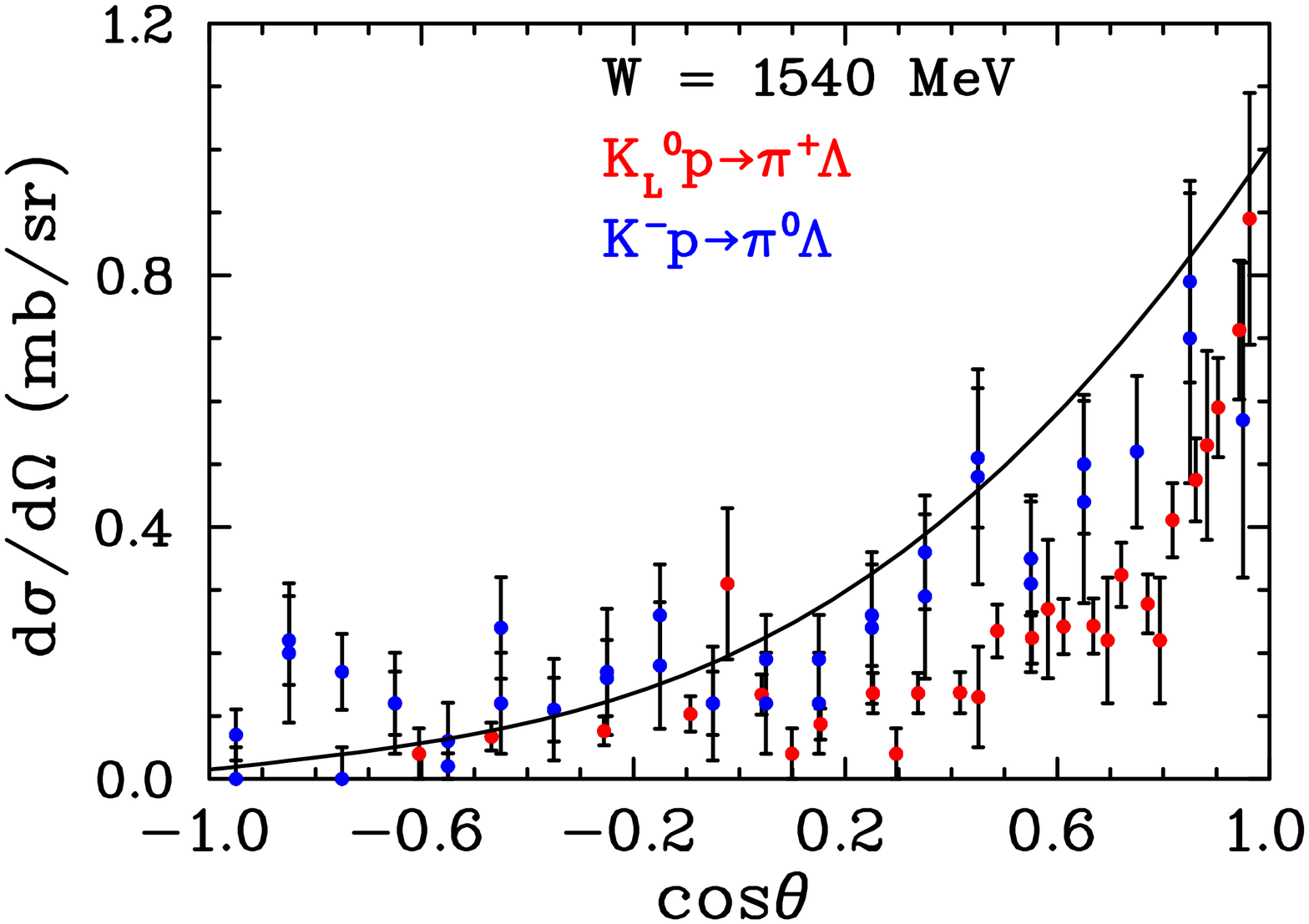}
\includegraphics[angle=90, width=0.35\textwidth]{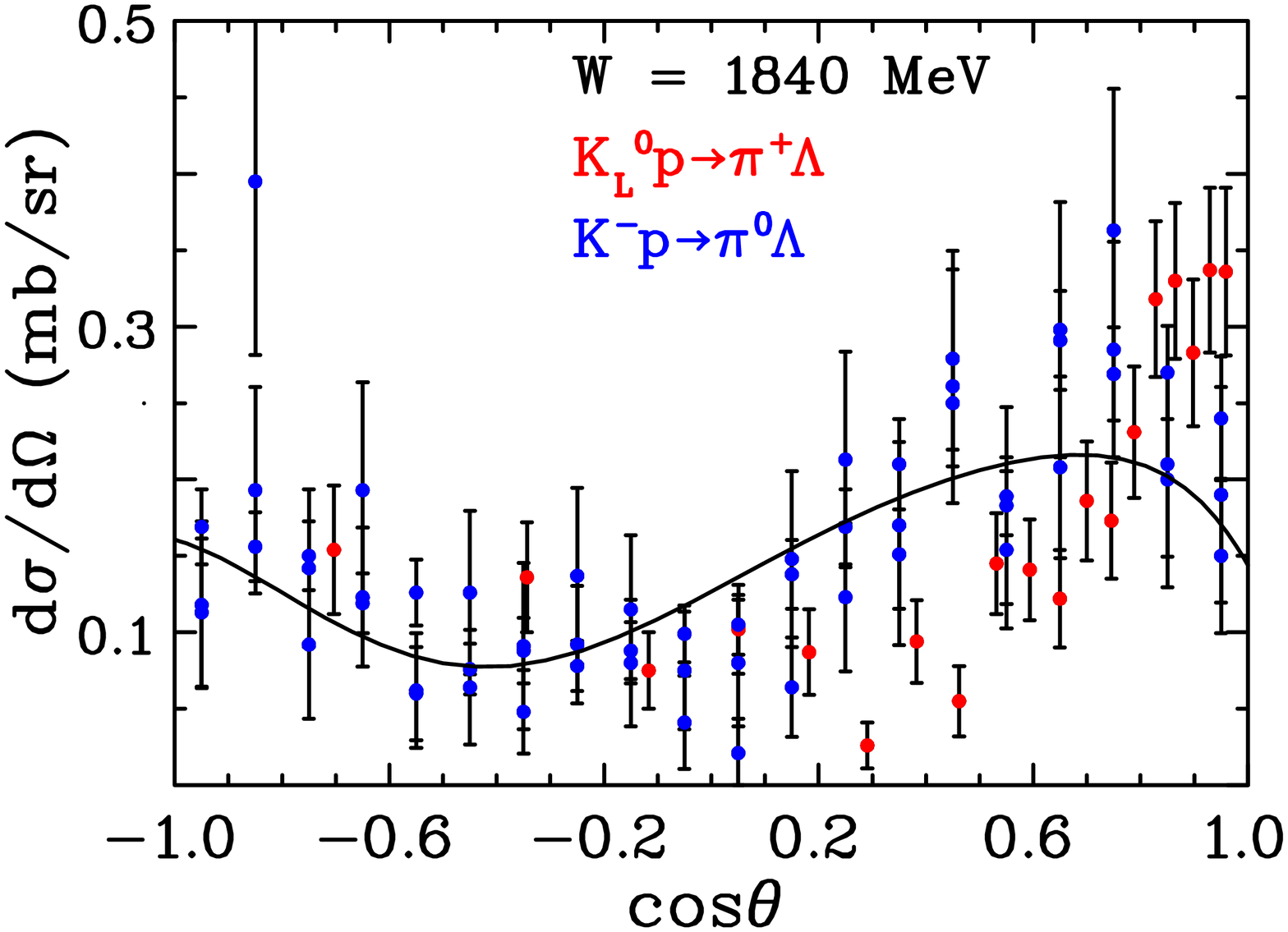}
\includegraphics[angle=90, width=0.35\textwidth]{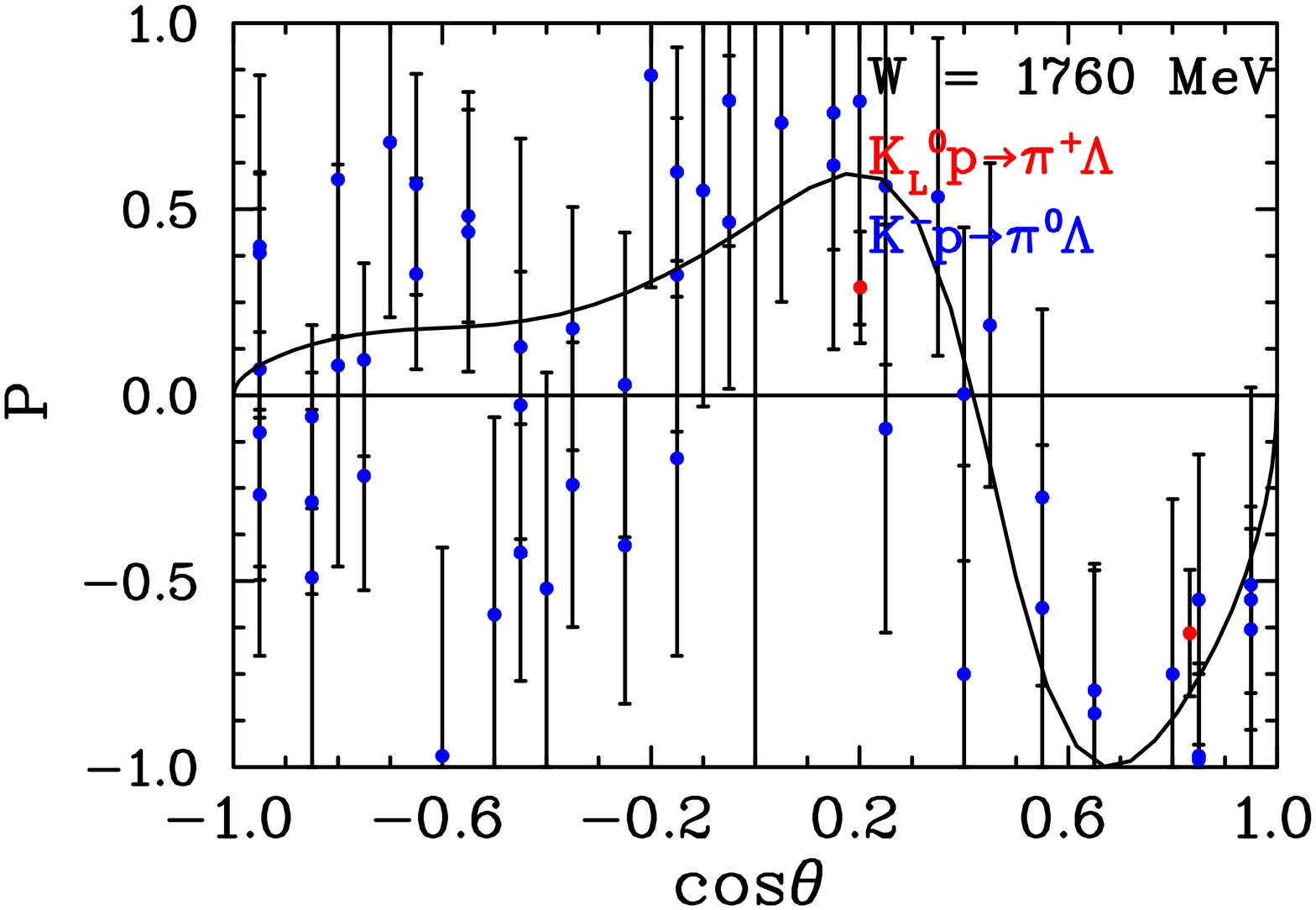}
\includegraphics[angle=90, width=0.35\textwidth]{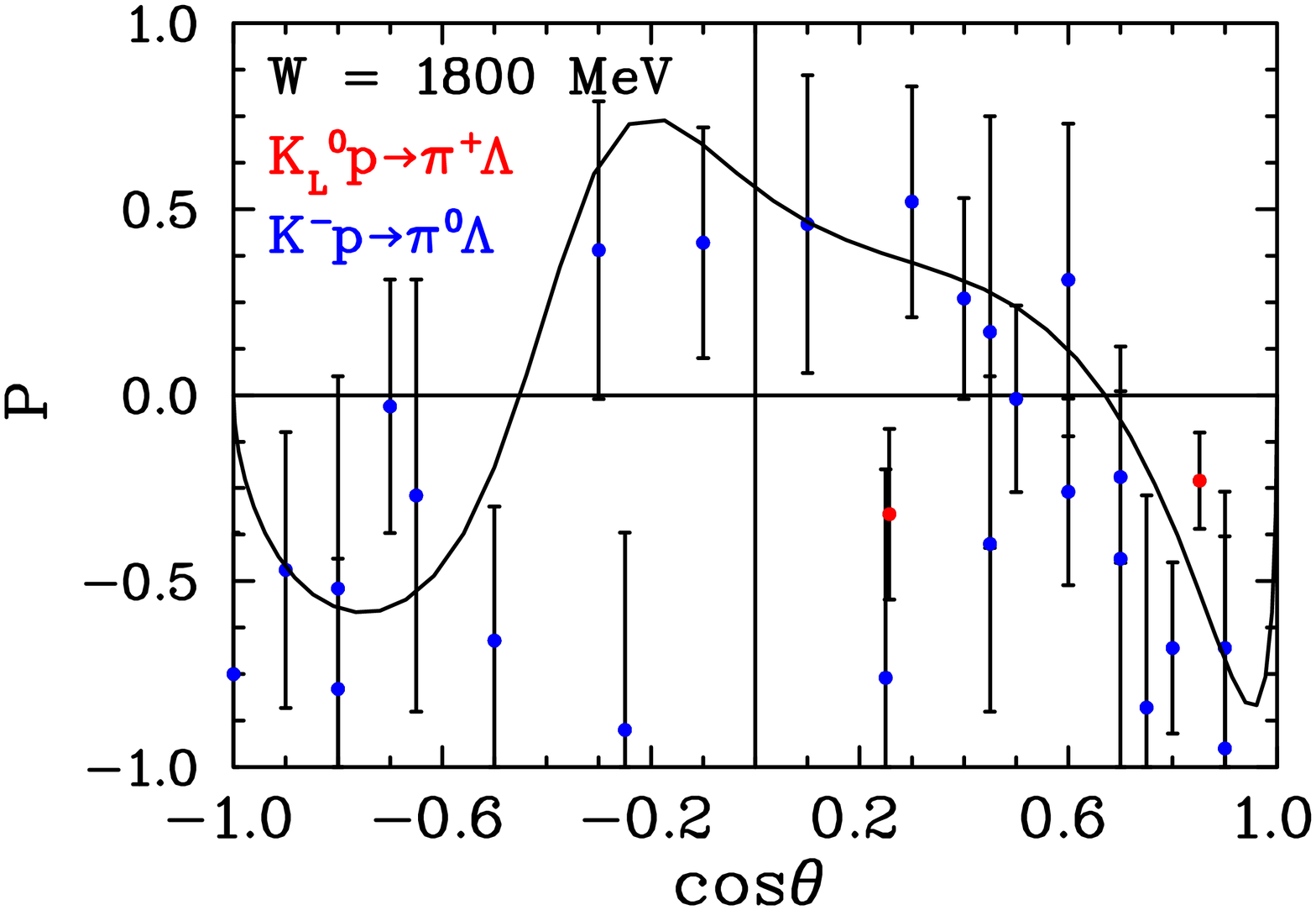}
\end{center}
\centerline{\parbox{0.80\textwidth}{
 \caption[] {\protect\small Comparison of selected differential
        cross section data for $K^-p\rightarrow\pi^0\Lambda$ and
        $K_Lp\rightarrow\pi^+\Lambda$ at W = 1540~MeV and 
	1840~MeV and selected polarization data at W = 1760~MeV
	and 1800~MeV from previous measurements are those 
	data points within 20~MeV of the Kaon c.m. energy 
	indicated on each panel~\protect\cite{Brem}.
        The curves are from the recent PWA of 
	$K^-p\rightarrow\pi^0\Lambda$ 
	data~\protect\cite{Zhang2013}.  }
        \label{fig:fig2} } }
\end{figure}

The $K^-p\rightarrow\pi^0\Lambda$ and $K_Lp\rightarrow\pi^+\Lambda$
amplitudes imply that observables for these reactions measured
at the same energy should be the same except for small
differences due to the isospin-violating mass differences in
the hadrons. No differential cross section data for $K^-p\rightarrow
\pi^0\Lambda$ are available at c.m.\ energies $W < 1540$~MeV,
although data for $K_Lp\rightarrow\pi^+\Lambda$ are available at
such energies due to longer $K_L$ life time.  At 1540~MeV and
higher energies, differential cross section and polarization
data for both reactions are in fair agreement, as shown in
Fig.~\ref{fig:fig2}.  Meanwhile, the quality of avilable P
measurements do not have a sensitivity to the fit.

\section{Proposed Measurements}
\label{Experiment}

We propose to use a Hall~D Facility with the GlueX spectrometer, 
to perform precision measurements of two-body reactions induced 
by the $K_L$-beam on the liquid hydrogen cryotarget in the 
resonance region, W = 1490 -- 4000~MeV and c.m. $\cos\theta$ 
from -0.95 to 0.95.  This ability of the GlueX provides an 
ideal environment for these experiment.
\begin{figure}[ht]
\begin{center}
\includegraphics[angle=0, width=0.9\textwidth ]{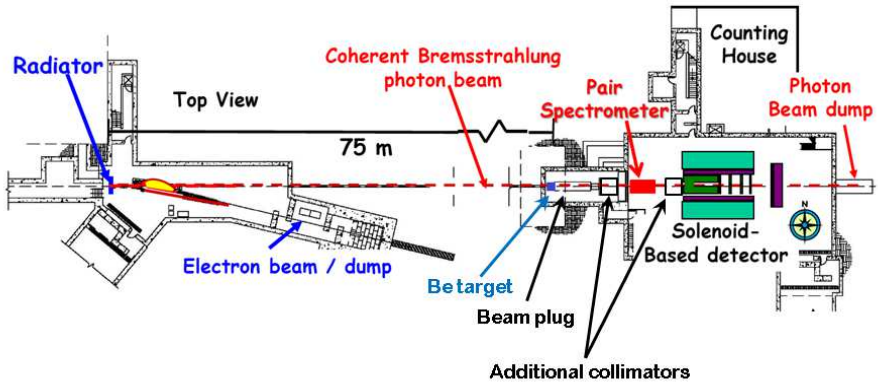}
\end{center}
\centerline{\parbox{0.80\textwidth}{
 \caption[] {\protect\small Schematic view of Hall~D
        beamline on the way $e\rightarrow\gamma\rightarrow K_L$.} 
	\label{fig:fig3}}}
\end{figure}

There is an advantage factor for $K_Lp$ vs. $K^-p$ experiment.
The mean lifetime of the $K_L$ is 51.16~ns ($c\tau = 15.3$~m) 
whereas the mean lifetime of the $K^-$ is 12.38~ns ($c\tau = 
3.7$~m)~\cite{PDG2014}.  For this reason, it is much easier to 
perform measurements of $K_Lp$ scattering at low beam energies 
compared with $K^-p$ scattering~\cite{Noumi2016}. 

The recently constructed GlueX detector in Hall-D is a large 
acceptance spectrometer with good coverage for both charged 
and neutral particles that can be adapted to this 
purpose~\cite{Curtis2016}.  Schematic view of the Hall~D 
beamline is presented in Fig.~\ref{fig:fig3}.  At the first 
stage, E$_e$ = 12~GeV electrons produced at the CEBAF will 
scatter in a radiator in the target vault, generating 
intensive beam of bremsstrahlung photons (we will not need 
in the Hall~D Broadband Tagging Hodoscope).  At the second 
stage, bremsstrahlung photons, created by electrons, hit 
the Be-target and produce $K_L$-mesons along with neutron 
and charged particles.  Finaly, $K_L$ will reach the LH$_2$ 
cryogenic target within GlueX settings.

\begin{figure}[ht]
\centerline{
\includegraphics[height=0.3\textwidth, angle=0]{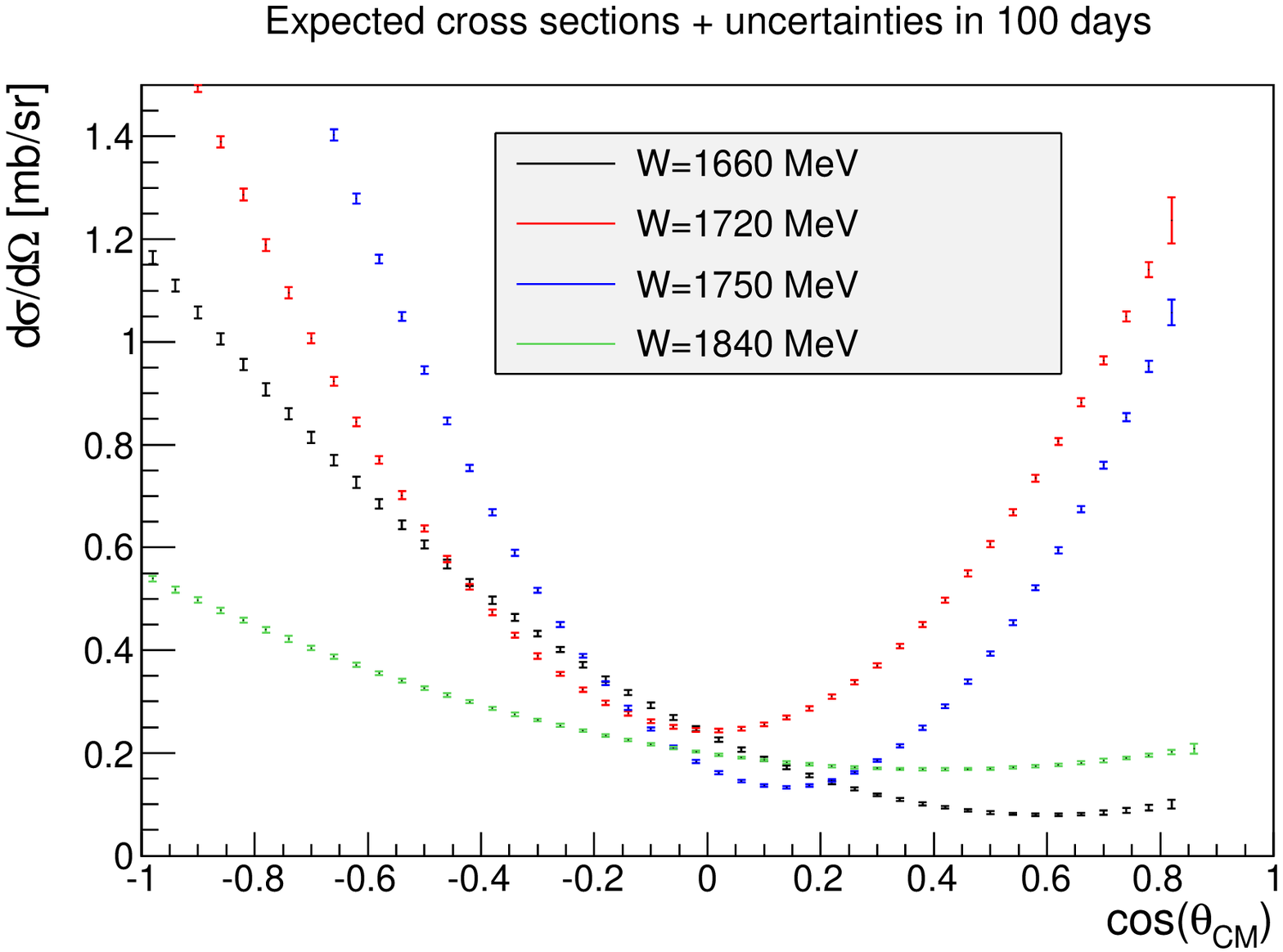}\\
\includegraphics[height=0.3\textwidth, angle=0]{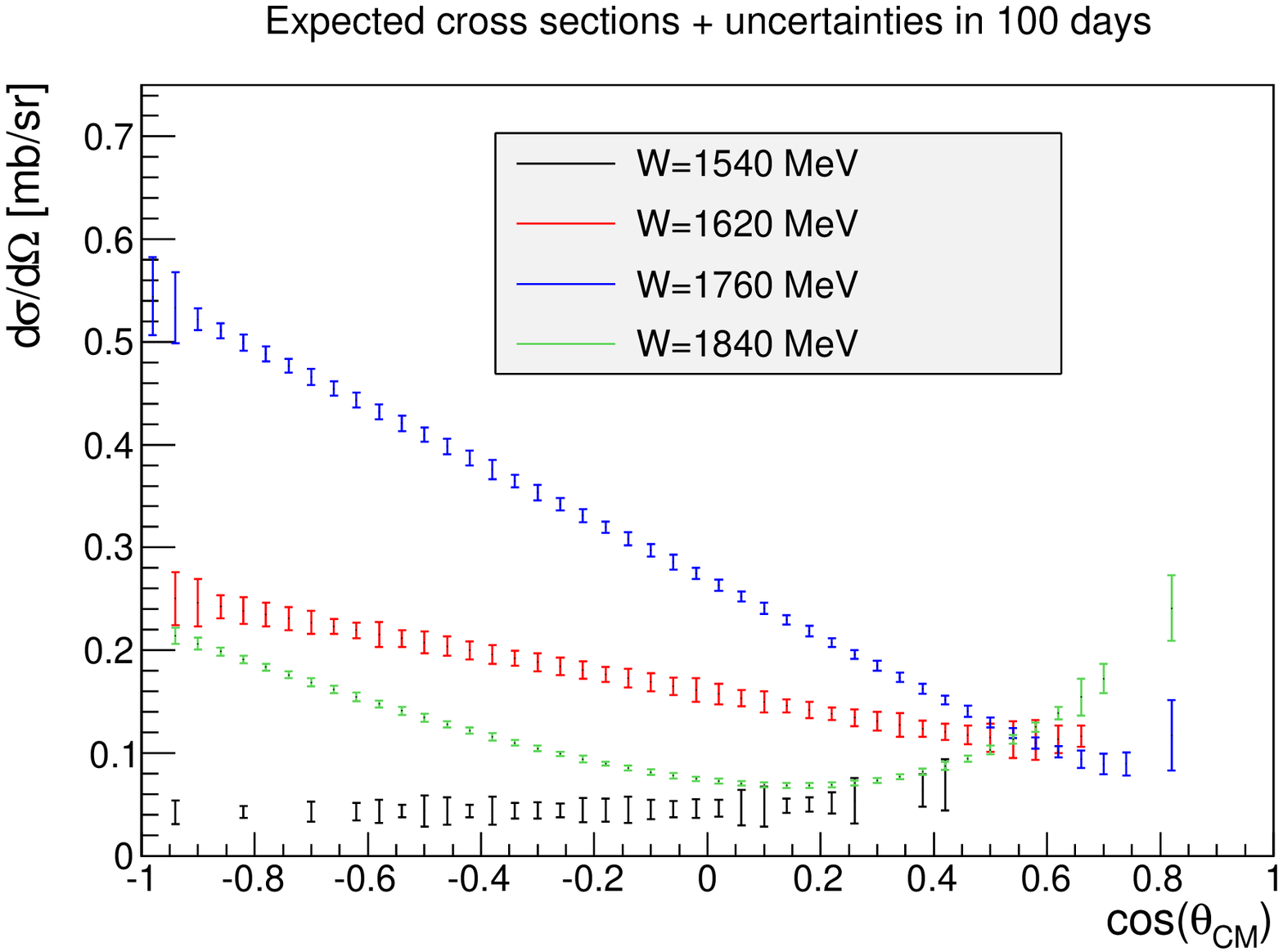}}
\caption{The cross section uncertainty estimates
        (statistics only) for $K_Lp\rightarrow pK_S$ (left)
        and for $K_Lp\rightarrow\pi^+\Lambda$ (right).
        \label{fig:fig5}}
\end{figure}

We estimated the flux of $K_L$ beam on the GlueX LH$_2$ 
target is about $10^5~K_L/s$, to be compared to about 
$10^3~K_L/s$ used at NINA~\cite{Albrow1970} and
SLAC~\cite{Brandenburg1973}, almost comparable
to charged Kaon rates obtained at AGS and elsewhere in
the past and expected for J-PARC~\cite{Noumi2016}. Momenta 
of neutral Kaons at JLab will be measured applying the 
time-of-flight technique using a time structure of 60~ns.  
The count rate estimates carried out assuming 100~days of 
data taking are presented in Fig.~\ref{fig:fig5}.

\section{Conclusion and Perspectives}
\label{Conclusion}

Precise new data (both differential cross section and recoil 
polarization of hyperons) for $K_Lp$ scattering with good 
kinematic coverage could significantly improve our knowledge on
$\Lambda^\ast$ and $\Sigma^\ast$ resonances.  Clearly, complete 
understanding of three-quark bound states requires to learn 
more about baryon resonances in "strange sector".  Polarization 
data are very important to be measured in addition to 
differential cross sections to help remove ambiguities in PWAs. 

Unfortunately, the current database for $K_Lp$ scattering 
includes very few polarization data. As noted here, several 
$K_Lp$ reactions are isospin (I=1) selective, which would 
provide a useful constraint for a combined PWA of $K_Lp$ and 
$K^-p$ reactions. Finally, the long lifetime of the $K_L$ 
compared with the $K^-$ would allow $K_Lp$ measurements 
to be made easier at lower energies compared with $K^-$ 
beams.  It would be advantageous to combine all $K_Lp$ data 
in a new coupled-channel PWA with available and new J-PARC 
$K^-p$ data when they will be available.  The proposed KL 
facility potentially may unravel many "missing" hyperons.
To complete SU(3)$_F$ multiplets, one needs no less than 17 
$\Lambda^\ast$, 41 $\Sigma^\ast$, 41 $\Xi^\ast$, and 24 
$\Omega^\ast$.

Measurements of "missing" hyperon states with their spin-parity  
assignments along with the "missing" non-strange baryons will 
provide very important ingredients to test QM and LQCD 
predictions thereby improving our understanding of QCD in a 
non-perturbative regime.\\

\begin{acknowledgement}
I thank Moskov Amaryan, Yakov Azimov, William Briscoe, Eugene
Chudakov, Ilya Larin, Mark Manley, James Ritman, and, Simon 
Taylor for comments on the feasibility of future measurements. 
This work is supported, in part, by the U.S.~Department of 
Energy, Office of Science, Office of Nuclear Physics, under 
Award Number DE--SC0014133.
\end{acknowledgement}


\end{document}